# Temperature dependent magnetoelectric response of lead-free $Na_{0.4}K_{0.1}Bi_{0.5}TiO_3$/$NiFe_2O_4$ laminated composites

Adityanarayan Pandey[a,*], Amritesh Kumar[b], Pravin Varade[a], K. Miriyala[a], A. Arockiarajan[b,c], Ajit. R. Kulkarni[a], N. Venkataramani[a,*]

[a]Department of Metallurgical Engineering and Materials Science, Indian Institute of Technology Bombay, Mumbai, 400076, India.

[b]Department of Applied Mechanics, Indian Institute of Technology Madras, Chennai, 600036, India.

[c]Ceramic Technologies Group-Center of Excellence in Materials and Manufacturing for Futuristic Mobility, Indian Institute of Technology Madras (IIT Madras), 600036 Chennai, India.

*Email: anbp.phy@gmail.com (A. Pandey), ramani@iitb.ac.in (N. Venkataramani)

**Abstract.** This study investigates the temperature-dependent quasi-static magnetoelectric (ME) response ($\alpha_E$) of electrically poled lead-free Na0.4K0.1Bi0.5TiO3-NiFe2O4 (NKBT-NFO) laminated composites. The aim is to understand the temperature stability of ME-based sensors and devices. The relaxor ferroelectric nature of NKBT is confirmed through impedance and polarization-electric (PE) hysteresis loop studies, with a depolarization temperature ($T_d$) of approximately 110°C. Heating causes a decrease and disappearance of planar electromechanical coupling ($K_p$), charge coefficient ($d_{31}$), and remnant polarization ($P_r$) above Td. The temperature rise also leads to a reduction in magnetostriction ($\lambda$) and magnetostriction coefficient ($q = d\lambda/dH$) of NFO by approximately 33% and 25%, respectively, up to approximately 125°C. At room temperature, the bilayer and trilayer configurations exhibit maximum ME responses of approximately 33 mV/cm·Oe and 80 mV/cm·Oe, respectively, under low magnetic field conditions (H ~ 300-450 Oe). The ME response of NKBT/NFO is highly sensitive to temperature, decreasing with heating in accordance with the individual temperature-dependent properties of NKBT and NFO. This study demonstrates a temperature window for the effective utilization of NKBT/NFO-based laminated composite ME devices.

## 1. Introduction

Magnetoelectric (ME) coupling between ferroelectric (FE) and ferro-/ferrimagnetic (FM) order parameters in multiferroic materials allows polarization to be controlled by magnetic field and magnetization by an electric field [1]. This opens up numerous applications such as, ultra-fast and low-power consuming ME Random access memory (MeRAM), telecommunication devices, ME-based sensors, spintronics, and energy harvesting devices [2]–[5]. However, single-phase ME materials are rare and most known systems have intrinsic drawbacks, such as low FE and/or FM ordering temperature which restricts application temperature and very weak ME effect at room temperature (RT) [6]. ME composites constituting FE and FM materials in different geometries (2-2: laminate, 0-3: particulate, and 1-3: rod/fiber composites, etc.) interacting with each other via elastic interactions, address the issue of single phase ME materials. These composites show large ME effect at RT at low bias magnetic field. Amongst different types of composites, 2-2 layered structures are easy to synthesize, easy to pole in order to get high piezoelectric responses from FE, exhibit large ME responses and exhibit reduced leakage currents [6]. The direct magnetoelectric effect (DME) effect is observed, when a magnetic field is applied to composite, the FM phase changes its dimension magnetostrictively, and the resultant strain is then transferred to the FE phase, resulting in an electric polarization. The DME response is often measured in voltage ($V$) (or electric field, $E$) induced in a composite due to an external ac magnetic field ($H$) as a function of DC magnetic field. The corresponding magnetic field induced ME voltage response ($\alpha_E$) is calculated using the following equation [7]

$$\alpha_E = \frac{dE}{dH} = \frac{1}{t}\left(\frac{dV}{dH}\right) \tag{1}$$

where $\alpha_E$ is the second-rank ME response tensor with the unit V/cm·Oe and '$t$' is the thickness of the FE layer in a laminated composite. The ME effect in bulk composites is extrinsic, depends on the microstructure, piezoelectric coefficient of FE layer, magnetostriction of FM layer and coupling interactions across the FE-FM interfaces. For laminated composites, the magnetoelectric voltage response ($\alpha_E$) is given as [7], [8]

$$\alpha_E = \frac{\delta E}{\delta H} = \frac{-2 d_{31}^p q_{11}^m v_m}{(S_{11}^m + S_{12}^m)\epsilon_p v_p + (S_{11}^p + S_{12}^p)\epsilon_p v_m - 2(d_{31}^p)^2 v_m} \tag{2}$$

where, $d_{31}$ is piezoelectric coefficient, $q_{11}$ =d$\lambda$/d$H$ is piezomagnetic coefficient ($\lambda$ is magnetostriction), $v_m$ and $v_p$ are volume fractions of the FM and FE phases, $\varepsilon$ is the dielectric constant, $S$ is the compliance coefficient. The compliance coefficient of most of the ceramics are similar. Therefore, the volume fraction ($v_m$ and $v_p$) of different layers and materials with higher individual properties ($d_{31}$ & $q_{11}$) are considered to enhance ME response of the laminated composite systems.

During the past two decades, mostly ME studies have been performed on composites of lead-based piezoelectric PZT (200-600 pm/V) or PMN-PT (1000-2500 pm/V) and magnetostrictive Terfenol-D ($Tb_{0.3}Dy_{1.7}Fe_2$: 1300-1400 ppm), due to their higher piezoelectric and magnetostrictive properties, respectively [6], [9]–[11]. Notably, the presence of lead in traditional piezoelectrics material raises environmental and health related concern. Additionally, and rare earth elements (Tb and Dy) based magnetostrictive materials Terfenol-D suffers from limitations of being highly brittle, anisotropic, and expensive. Hence, we have attempted to investigate the magnetoelectric response in a 2-2 laminated composite of lead-free $Na_{0.4}K_{0.1}Bi_{0.5}TiO_3$ (NKBT) and $NiFe_2O_4$ (NFO). NKBT is a morphotropic phase boundary (MPB) composition exhibits good dielectric, ferroelectric, piezoelectric properties with high planar electromechanical coupling [12]–[16]. NFO is a suitable magnetic phase exhibiting good magneto-mechanical coupling, low coercivity, moderate magnetostriction and permeability [6], [17], [18]. Recently, we have reported room temperature magnetoelectric response in bilayer and trilayer NKBT/NFO bulk composites which gave larger ME response in comparison to other lead-free FEs and ferrite bulk laminated systems [19].

From industrial and device point of view, ME based sensors and other devices are operated in demanding environmental conditions including high temperature range. There are only a few reports on composites (PZT/Metglas, PZT/Ni, PZT/Terfenol-D, PZT/LSMO, PVDF/transition metal based alloy, $0.37BiScO_3$-$0.63PbTiO_3$ (BS-PT)/NFO, etc.) which show the temperature effect on ME coupling in laminated composites till date [20]–[29]. In the present study, individual dielectric, ferroelectric and electromechanical coupling factor of NKBT and magnetostrictive behavior of NFO have been obtained at different temperatures from 27 °C to 125 °C. This has been then correlated with temperature dependent quasi-static ME behavior of the synthesized layered composite samples.

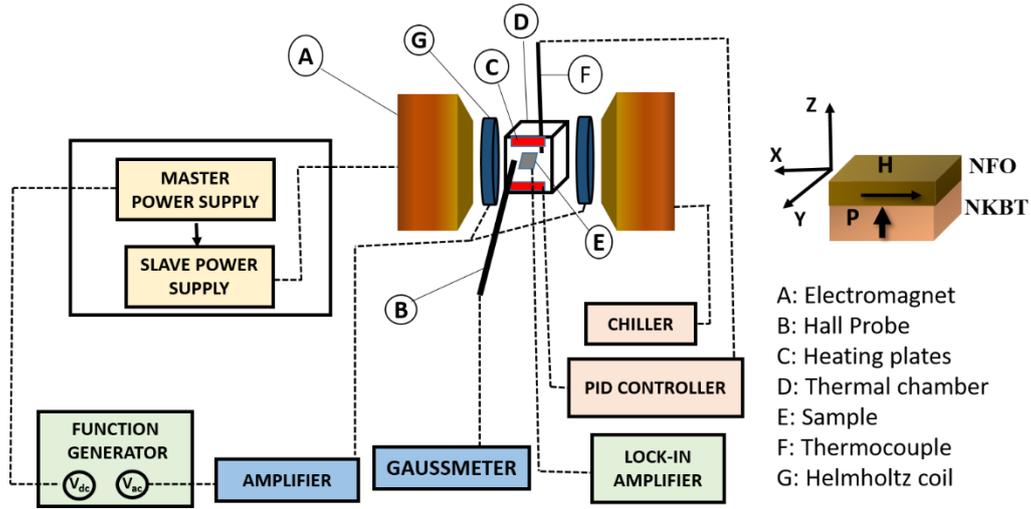

**Figure 1.** Schematic diagram of the experimental set-up for measuring temperature and magnetic field dependent ME response ($\alpha_E$).

## 2. Experimental

Ceramic samples of NKBT and NFO are synthesized and sintered as described in ref [19], [30]. The density of each rectangular pellet measured using the Archimedes method after sintering was ≥97% of the theoretical density. Sintered NKBT and NFO ceramics of 1 mm and 2 mm thicknesses are individually machined to an area of 7x7.5 mm$^2$ and both the pellets with bilayer (NKBT/NFO) and trilayer configurations of (NFO/NKBT/NFO) are electroded using silver paste. Later, the laminate composites are sandwiched by bonding each NKBT and NFO using a thin layer plate conductive silver epoxy with resin to a hardener ratio of 1:1 and cured at 100 °C for 4 h. The composites are electrically poled (polarized) perpendicular to its plane before the ME measurement. Electrical poling is performed by heating the sample to 120 °C and cooling back to RT (27 °C) under a constant applied voltage of 28 kV/cm for 1h. The alpha-A high-resolution impedance analyzer (Novocontrol GmbH, Germany) is used to measure the impedance and dielectric response of ceramics in the frequency range of 1 Hz to 1 MHz. Various parameters of NKBT, viz., planar electromechanical coupling ($K_p$), transverse electromechanical coupling ($K_{31}$), elastic compliance ($S_{11}$), charge coefficient ($d_{31}$), and voltage coefficients ($g_{31}$) etc. are calculated using resonance ($f_r$) and anti-resonance ($f_a$) method following IEEE standards given as [31], [32]

$$\frac{1}{K_p^2} = \frac{0.395 f_r}{f_a - f_r} + 0.574; \quad \frac{1}{S_{11}} = \frac{\pi^2 \varphi^2 f_r^2 (1-\sigma^2)\rho}{\eta^2}; \quad g_{31} = \frac{d_{31}}{\epsilon_{33}};$$

$$K_{31}^2 = \left(\frac{1-\sigma}{2}\right) K_p^2; \text{ and } d_{31} = K_{31}\sqrt{\epsilon_{33} S_{11}} \tag{3}$$

where $\sigma$ = 0.3 is poisson's ratio, $\varphi$ is diameter and $\rho$ is density of rectangular ceramic pellet, $\eta$ =2.05 is a dimensionless constant, $\varepsilon_{33}$ is dielectric constant. Field-induced polarization measurements are performed at 50 Hz using a standard ferroelectric PE loop tester (aixACCT GmbH, TF 2000). Magnetostriction measurements have been conducted on the synthesized samples of dimension (7x7.5x2 mm$^3$) using strain gauge connected to the strain indicator (Syscon-5CH). The sample is placed in a specially designed stainless steel (Grade 304) thermal chamber, wherein temperature is maintained by a PID controller. The chamber is placed in between the poles of an electromagnet (GMW 5403) and the magnetic field near the sample is measured by the use of a hall probe in combination with a gaussmetre (Lakeshore F41). For ME measurements, in addition to the above mentioned instruments, a lock-in amplifier (SRS-865A) is used in place of strain indicator. Additionally, in combination of DC bias field, the required AC field at a constant operating frequency of 1 kHz is provided using a pair of Helmholtz coil receiving power from an amplifier (KEPCO, BOP 36-6DL). The DC and the AC signal are generated by a function generator (Tektronix AFG3022C). The schematic diagram of the experimental set-up for measuring ME response as a function of magnetic field and temperature is shown in Fig. 1.

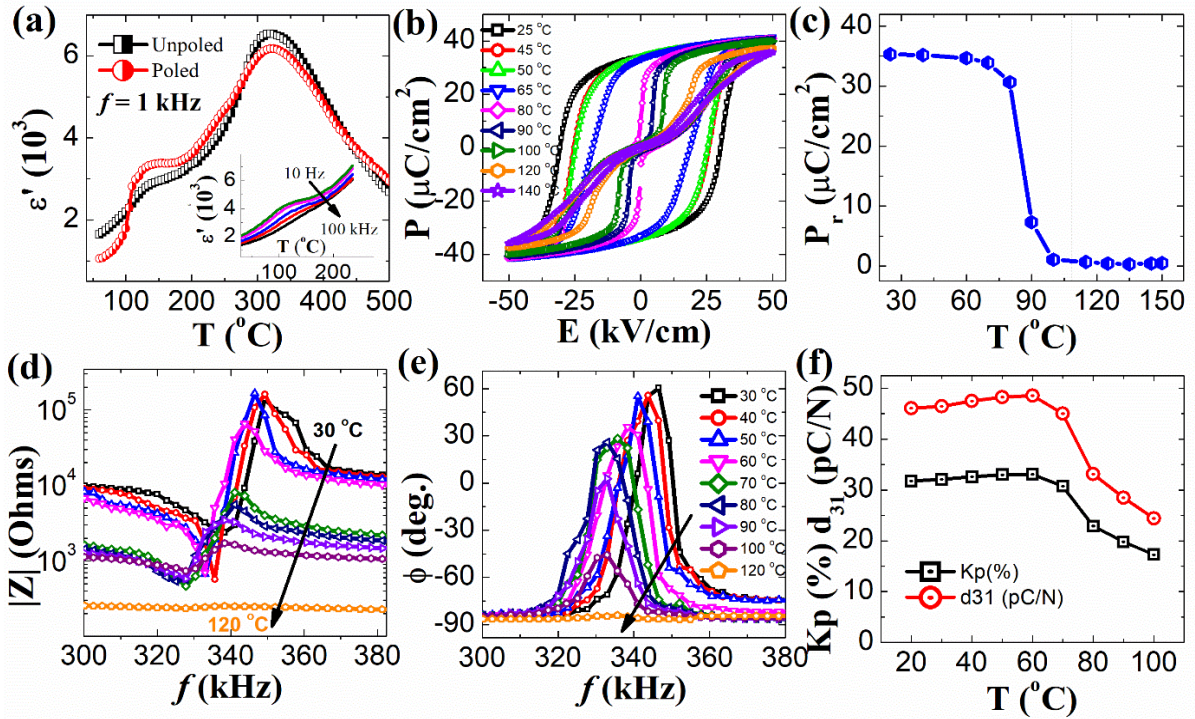

**Figure 2.** (a) The variation of dielectric constant ($\varepsilon'$) as a function of temperature at 1 kHz frequency for poled and unpoled NKBT ceramic sample; inset of Fig. 1(a) shows variation of loss tangent (tan $\delta$) as function of temperature for poled/unpoled cases depicting depolarization temperature ($T_d$), (b) P-E hysteresis loop and, (c) Remnant polarization ($P_r$) as function of temperature at 50 Hz frequency, the variation of (d) Impedance (|Z|) and (e) Phase ($\phi$) as function of frequency (f) for poled ceramic sample at different temperatures, and (f) The variation of planar electromechanical coupling ($K_p$) and charge coefficient ($d_{31}$) as a function of temperature.

## 3. Results and discussion

Figure 2(a) shows the temperature dependent dielectric constant ($\varepsilon'$) for unpoled and poled NKBT ceramic sample at 1 kHz frequency. It reveals two broad dielectric peaks in the measured temperature range of 30-500 °C; the first hump near ~110 °C represents relaxor behavior [Inset Fig 2(a)], and the second high-temperature broad peak signifies a diffuse FE phase transition temperature $T_c$ ~310 °C [33]. A distinct peak in the loss tangent of poled ceramic (not shown here), corresponds to the depolarization temperature ($T_d$), where a step-like increase in the $\varepsilon'$ is also evident near $T_d$. Figure 2(b) shows the P-E loops at different temperatures from 30 °C to 150 °C measured for NKBT ceramics under an applied electric field of 50 kV/cm. A typical ferroelectric hysteresis P-E loop is observed at room temperature (RT), with high remnant polarization ($P_r$) ~35 μC/cm$^2$ and coercive field ($E_c$) ~30 kV/cm. Shape of P-E loop changes from a typical rectangular shape to a pinched loop with increase in temperature and the value of $P_r$ significantly decreases above 70 °C and becomes approximately zero near $T_d$ [Fig. 2(c)]. This is consistent with the low temperature dielectric anomaly in poled sample. Figure 2(d,e) shows a typical frequency dependent impedance (|Z|) and phase ($\phi$) for poled ceramic sample at different temperatures. The resonance (lowest impedance, $f_r$) and anti-resonance (highest impedance, $f_a$) frequencies are observed in the frequency range of 300-380 kHz. An ideal piezoelectric sample exhibit 180° phase shift (from -90° to +90°) with increasing frequency from $f_r$ to $f_a$ implying complete poling. The observed phase shift ~150° at room temperature for NKBT is less in comparison to ideal value which may be associated with the internal stresses and field-induced lattice distortion impeding domain switching as suggested by Li et al. [34]. With increase in temperature, a significant reduction in phase shift is observed and it completely disappears after 110 °C implying depolarization of NKBT ceramic sample. Further, the planar electromechanical coupling ($K_p$) and charge coefficient ($d_{31}$) are calculated using impedance data following IEEE standards [31], [32]. Figure 2(f) shows that the value of $K_p$ decreases from 32 to 18 (%) and $d_{31}$ decreases from 48 to 26 pC/N with rise in temperature and both disappear above $T_d$ due to depolarization NKBT akin to $P_r$ shown earlier. The values of $K_p$ and $d_{31}$ are close to recently reported NKBT(82/18) ceramic [35], [36].

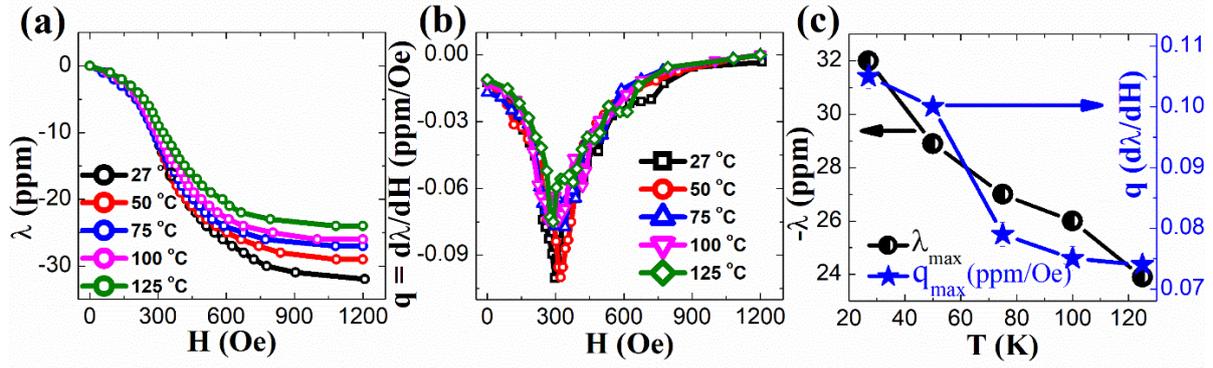

**Figure 3.** The variation of (a) magnetostriction ($\lambda$) and (b) magnetostriction coefficient ($q = d\lambda/dH$) as function of dc-magnetic field ($H$) at different temperatures 27-125 °C, and (c) The variation of $\lambda_{max}$ and $q_{max}$ as function of temperature for NFO ceramic sample.

Figure 3(a,b) shows the magnetic field ($H$) dependence of magnetostriction ($\lambda$) and magnetostriction coefficient ($q = d\lambda/dH$) at different temperatures from RT to 125 °C for NFO ceramic sample, respectively. The $\lambda$ is negative owing to the negative magnetostrictive nature of NFO and its value increases with increasing $H$ and then saturates for $H \geq 1$ kOe for all temperatures. The value of $q$ also shows maximum when $\lambda$ undergoes maximum change in slope ~300-450 Oe and then it saturates for $H \geq 1$ kOe for all temperatures. The RT value of $\lambda_{sat}$ ~32 ppm and $q_{max}$ ~0.105 ppm/Oe match well with the values reported in literature for NFO [18]. Noticeably, both $\lambda_{sat}$ and $q_{max}$ decrease with increasing temperature. At 125 °C, the drop in $\lambda_{sat}$ and $q_{max}$ is 33% and 25%, respectively w.r.t. RT [Fig. 3(c)]. It is known that the magnetization and the magnetostriction of magnetic materials are correlated as: $\varphi \sim 3\lambda\sigma/(K+2\pi M^2)$, i.e., the $\lambda$ is proportional to $M^2$; where $\varphi$ is the angle between the magnetization and applied field, $\sigma$ is stress, and $K$ is the anisotropy constant [37], [38]. NFO is a ferrimagnetic ferrite material with a high Curie temperature $T_c$~570 °C and reduction in magnetostriction on increasing temperature is consistent with magnetization behavior [39]. Further, the temperature dependent magnetoelectric response ($\alpha_E$) is measured as a function of magnetic field and corroborated with piezoelectric and magnetostriction behavior of NKBT and NFO, respectively.

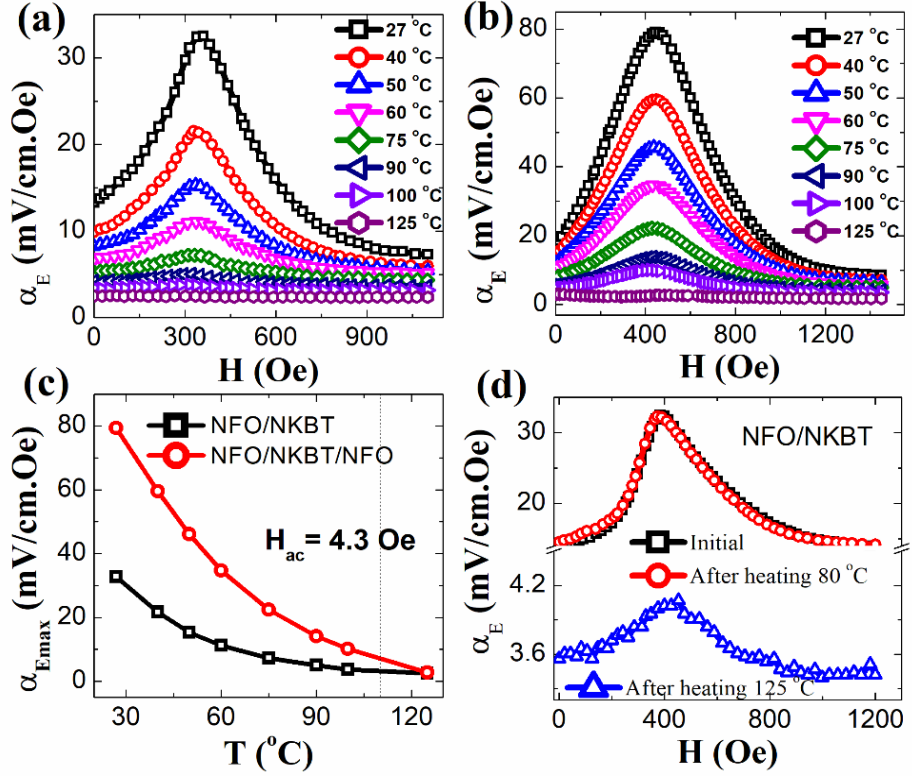

**Figure 4.** The variation of $\alpha_E$ as function of DC-magnetic field ($H$) at 1 kHz frequency and an applied field of 4.3 Oe ac magnetic field measured at different temperatures 27-125 °C for (a) NKBT/NFO (bilayer), (b) NFO/NKBT/NFO (trilayer) laminated composites, (c) The variation of maximum ME response ($\alpha_{Emax}$) as a function of temperature, and (d) The $\alpha_E$ as function of H of bilayer composite sample measured at room temperature after different thermal history.

Figure 4(a,b) shows variation of $\alpha_E$ as function of DC-magnetic field ($H$) for NKBT/NFO (bilayer) and NFO/NKBT/NFO (trilayer) laminated composites, respectively, at 1 kHz frequency and an applied AC-field of 4.3 Oe at temperatures ranging from 27-125 °C. When $H$ is varied, the generated mechanical strain in NFO layer is transferred to piezoelectric NKBT layer through silver epoxy, and the generated ME voltage ($V$) across NKBT is measured using a lock-in amplifier. At room temperature, ME response first increases and reaches a maximum of 33 and 80 mV/cm.Oe at $H$~300-450 Oe, for bilayer and trilayer, respectively. Eventually, the $\alpha_E$ decreases with the dc-field and saturates above $H \geq 1$ kOe for both samples. The ME response with $H$ can be explained considering Eq. 2, as $\alpha_E$ is proportional to magnetostriction coefficient of NFO ($\alpha_E = dE/dH = dE/d\lambda \cdot d\lambda/dH$; where $q = d\lambda/dH$) and the maximum $\alpha_E$ is associated with the maximum change in $\lambda$ of NFO layer with an applied field $H$. With increasing temperature $\alpha_E$ curve becomes flat after $T_d$ which is in good corroboration with $P_r$ and $d_{31}$ plots of NKBT. Figure 4(c) also shows that maximum ME response ($\alpha_{Emax}$) decreases with increasing temperature and $\alpha_{Emax}$ becomes nearly zero at 125 °C for both samples. This implies that ME response of NKBT/NFO bonded by is highly dependent on temperature. The rapid decrease in $\alpha_{Emax}$

with temperature is related to decrease in the magnetostriction and piezoelectric properties of individual layers NFO and NKBT, respectively, which is consistent with earlier literatures for different laminated composites [22], [24], [28], [29], [40]. Apart from temperature effect of $\lambda$ and $d_{31}$, the drop in $\alpha_{Emax}$ with temperature is also related to a bonding medium i.e., Ag epoxy as suggested by Kumar *et al.* for BSPT/NFO composites [22]. They observe a large drop (~75%) in ME response with temperature for Ag epoxy-bonded BSPT/NFO laminated composites, in comparison to a small drop (~19%) in epoxy-free (co-sintered ceramic) case. Similar, temperature dependent ME response behaviour is also reported by Amritesh *et al.* for epoxy-free and epoxy-bonded PZT/Ni and PZT/FeNi layered composites [23], [24]. Further, the operating temperature range is checked by (i) heating composite sample to 80 °C (below $T_d$) and cooling back to RT and (ii) heating composite sample to 125 °C (above $T_d$) and cooling back to RT. The measured ME response for two cases are compared in Fig. 3(d). It suggests that if composite sample is heated below $T_d$, there is no change in $\alpha_E$ measured at RT. In contrast, the drop in $\alpha_E$ is ~88% for other case is due to depolarization of the piezoelectric layer as expected. The operating temperature range can be enhanced by increasing the depolarization temperature with further substitution in NKBT.

## 4. Conclusion

In summary, this study focused on investigating the temperature-dependent quasi-static magnetoelectric (ME) response of bilayer and trilayer lead-free $Na_{0.4}K_{0.1}Bi_{0.5}TiO_3$-$NiFe_2O_4$ (NKBT-NFO) laminated composites. The aim was to understand the temperature stability of ME-based sensors and devices. The results revealed that the ME response of the composites was highly sensitive to temperature, with a decrease in the response observed with heating. The individual temperature-dependent properties of NKBT and NFO, such as the depolarization temperature, planar electromechanical coupling, charge coefficient, magnetostriction, and magnetostriction coefficient, influenced the overall ME response of the laminated composite. The study demonstrated a temperature window for the effective utilization of NKBT/NFO-based laminated composite ME devices. These findings contribute to the understanding of temperature effects on ME coupling in lead-free laminated composites and have implications for the design and optimization of ME-based devices operating in demanding environmental conditions.

**Author contributions:**

All authors contributed to the study conception and design. Material preparation, data collection and analysis were performed by Adityanarayan Pandeya, Amritesh Kumar, Pravin Varade, K. Miriyala. The first draft of the manuscript was written by Adityanarayan Pandey and all authors commented on previous versions of the manuscript. All authors read and approved the final manuscript.

**Acknowledgements**

The authors acknowledge the IRCC, IIT Bombay, for dielectric, impedance and PE loop measurements. NV & ARK acknowledge Department of Science and Technology, India (Project Code No. RD/0118-DST000-020) for supporting this work. AA would like to extend his gratitude to Science and Engineering Research Board of India, DST, India for providing the financial aid under Project No.: EMR/2015/001559. The funding received from the Institute of Eminence Research Initiative Project on Materials and manufacturing for Futuristic mobility (Project no. SB20210850MMMHRD008275) is gratefully acknowledged. AP acknowledges Indian Institute of Bombay, Mumbai for the post-doctoral research fellowship.